\documentclass[aps,prd,twocolumn,superscriptaddress,groupedaddress]{revtex4} %,showpacs , showkeys
\usepackage{subfig, caption}
\usepackage{natbib}
\usepackage{mathtools}
\usepackage{txfonts}
\usepackage{scalerel}
\usepackage{graphicx}  % needed for figures
\usepackage{bm}        % for math
\usepackage{amssymb}   % for math
\usepackage{lipsum}
\usepackage{lineno}
\usepackage{amsmath,array}
\usepackage{caption}
\usepackage{hyperref}
\hypersetup{
	colorlinks=true,
	linkcolor=blue,
	filecolor=magneta,      
	urlcolor=blue,
}
\DeclareCaptionJustification{justified}{\leftskip=0pt \rightskip=0pt \parfillskip=0pt plus 1fil}

\usepackage{tikz}
\usetikzlibrary{svg.path}
\definecolor{orcidlogocol}{HTML}{A6CE39}
\tikzset{
    orcidlogo/.pic={
        \fill[orcidlogocol] svg{M256,128c0,70.7-57.3,128-128,128C57.3,256,0,198.7,0,128C0,57.3,57.3,0,128,0C198.7,0,256,57.3,256,128z};
        \fill[white] svg{M86.3,186.2H70.9V79.1h15.4v48.4V186.2z}
        svg{M108.9,79.1h41.6c39.6,0,57,28.3,57,53.6c0,27.5-21.5,53.6-56.8,53.6h-41.8V79.1z M124.3,172.4h24.5c34.9,0,42.9-26.5,42.9-39.7c0-21.5-13.7-39.7-43.7-39.7h-23.7V172.4z}
        svg{M88.7,56.8c0,5.5-4.5,10.1-10.1,10.1c-5.6,0-10.1-4.6-10.1-10.1c0-5.6,4.5-10.1,10.1-10.1C84.2,46.7,88.7,51.3,88.7,56.8z};
    }
}
\newcommand\orcidicon[1]{\href{https://orcid.org/#1}{\mbox{\scalerel*{
                \begin{tikzpicture}[yscale=-1,transform shape]
                \pic{orcidlogo};
                \end{tikzpicture}
            }{|}}}}
\begin{document}
    
%===============================================

\title{Constraining spinning primordial black holes with global 21-cm signal}
\author{Pravin Kumar Natwariya $^{\orcidicon{0000-0001-9072-8430}}$\,}
\email{pvn.sps@gmail.com, pravin@prl.res.in}
\affiliation{%
    Physical Research Laboratory, Theoretical Physics Division, Ahmedabad, Gujarat 380 009, India}
\affiliation{%
    Department of Physics, Indian Institute of Technology, Gandhinagar, Palaj, Gujarat 382 355, India}  
\author{Alekha C. Nayak $^{\orcidicon{0000-0001-6087-2490}}$\,}
\email{alekhanayak@nitm.ac.in}
\affiliation{%
    National Institute of Technology, Meghalaya, Shillong, Meghalaya 793 003, India}
\author{Tripurari Srivastava $^{\orcidicon{0000-0001-6856-9517}}$\,}
\email{tripurarisri022@gmail.com}
\affiliation{%
    Department of Physics and Astrophysics, University of Delhi, Delhi 110007, India}
\date{\today}
%===============================================
\begin{abstract}
\centerline{\textbf{\small Abstract}}

\vspace{0.1cm}
We study the upper  projected bounds on the dark matter fraction in the form of the primordial black holes (PBHs) with a non-zero spin by using the absorption feature in the global 21-cm signal at redshift $z \approx 17$. The mass and spin are fundamental properties of a black hole, and they can substantially affect the evaporation rate of the black hole. The evaporating black hole can inject energy into the intergalactic medium and heat the gas. Subsequently, it can modify the absorption amplitude in the global 21-cm signal. Therefore, the absorption feature in the 21-cm signal can provide a robust bound on PBHs. We analyse the  projected constraints on the dark matter fraction in the form of both spinning and non-spinning PBHs. The constraints are more stringent for spinning PBHs than non-spinning ones. We also compare these bounds with other observations and find the most stringent lower constraint on PBHs mass, which is allowed to constitute the entire dark matter to $6.7\times10^{17}$~g for extremal spinning PBHs.

%Primordial Black Holes can be a possible dark matter candidate. PBHs can account for as a fraction or all dark matter component in the Universe.

\end{abstract}
%===============================================
%\pacs{}
\maketitle
%===============================================
%===============================================
\section{Introduction}

About 85 percent of the total matter content in the Universe is dominated by the dark matter (DM) \cite{Planck:2018}. In the last decade, many DM models, such as collision-less cold DM  \cite{Peebles:1982}, fuzzy cold DM \cite{Hu:2000}, warm DM \cite{Dodelson:1994, Boyarsky:2019, Bulbul:2014}, self-interacting DM  \cite{Spergel:2000DSP, Natwariya:2020V}, have been proposed to explain various astrophysical observations. However, the microscopic nature of dark matter is still unknown. One interesting, well-motivated proposal is the fraction/all of DM in the form of PBHs (\cite{Carr:2020, Dasgupta:2020, Frampton:2010, Khlopov_2010, Belotsky_2019} and references therein).  Recently, PBHs have been gathered much attention after the black hole binary merger detection by LIGO and Virgo collaboration. These events suggest that PBHs may constitute a fraction of DM \cite{Bird:2016, Abbott:2016A, Abbott:2016B, Abbott:2016C, Abbott:2016D, Sasaki:2016}.

PBHs may have originated in the early Universe due to  initial inhomogeneities \cite{Zeldovich:1967, Hawking:1971, Carr:1974H, Carr:1975}, Higgs potential instability at a scale of $10^{11}$~GeV \cite{Espinosa:2018}, hybrid inflation \cite{Frampton:2010, Sebastien:2015}, etc. Depending on the origin time (t), PBHs can have a wide range of masses $M_{\rm PBH}\sim 10^{15}\,\big[t/(10^{-23}\rm sec)\big]$~g \cite{Carr:2010}. PBHs having mass larger than $10^{15}$~g can survive the Hawking evaporation and account for present-day DM density \cite{Hawking:1975}. The presence of PBHs can be  responsible for the  ultra-luminous X-ray sources, seeds for supermassive black holes at the centre of galaxies \cite{Sebastien:2015}, and it may provide seeds for present-day observed structures \cite{Sebastien:2015, Garc:2017}. There are several hints that indicate the presence of PBHs, such as dynamics and star clusters of ultra-faint-dwarf-galaxies, correlations between X-ray and infrared cosmic backgrounds, etc. (for a detailed review, see Ref. \cite{Sebastien:2018}). The presence of evaporating PBHs can explain the galactic/extra-galactic $\gamma$-ray background radiation \cite{Wright:1996, Lehoucq:2009, Carr:1976, Page:1976}, short-duration $\gamma$-ray bursts \cite{Cline:1997, Green:2001}, and reionization by injection of $\gamma$ and $e^{\pm}$ radiations into Inter-Galactic-Medium (IGM) \cite{Belotsky_2014, Belotsky:2015}.

During the cosmic dawn era, the evolution of the gas temperature and ionization fraction of the Universe are well-known \cite{Seager1999, Seager}. The addition of any exotic source of energy during the cosmic dawn era can significantly impact the ionization and thermal history of the Universe. Therefore, we can constrain the properties of such exotic sources from the observations during the cosmic dawn era. Evaporating PBHs can heat the gas and modify the free electron fraction in the IGM \cite{Laha:2021, Kim_2021}. Rotating PBHs can emit more particles into IGM and substantially affect the IGM evolution compared to non-rotating PBHs  \cite{Chandrasekhar:1977, Page:1976, Page::1976}. Therefore, it is important to study the properties of spinning PBHs. In the present work, we consider the Hawking emission of PBHs into background radiations (photons and electron/positron) and  provide the projected constraints on the fraction of DM in the form of PBHs, $f_{\rm PBH}=\Omega_{\rm PBH}/\Omega_{\rm DM}$, as a function of mass and spin. Here, $\Omega_{\rm PBH}$ and $\Omega_{\rm DM}$ are the dimensionless density parameters for PBHs and DM.

Recently, EDGES observation detected a large absorption signal in the 21-cm line in the redshift range $15-20$ \cite{Bowman:2018yin, Pritchard_2012}. The 21-cm signal appears to be a  treasure trove to provide constraints on various cosmological phenomena such as the formation of the first stars, galaxies or any exotic source of energy injection. The 21 cm line corresponds to the wavelength of the hyperfine transition between the $1S$ singlet and triplet states of the neutral hydrogen atom. The EDGES absorption signal is nearly two times larger than the theoretical prediction based on the standard model of cosmology ($\Lambda$CDM)  at the redshift $z\approx17$ \cite{Bowman:2018yin, Pritchard_2012}. In the  $\Lambda$CDM model, the gas and cosmic-microwave-background (CMB) temperatures vary adiabatically as $T_{\rm gas}\propto(1+z)^2$ and $T_{\rm CMB}\propto(1+z)$ during the cosmic dawn era. Subsequently,  at $z=17$, one gets the gas and CMB temperatures to be $\sim  6.7$~K and $\sim 49.1$~K, respectively, and it implies $T_{21}\approx-220$~mK \cite{Seager1999, Seager}. While EDGES collaboration reported  $T_{21}=-0.5_{-0.5}^{+0.2}$~K with 99\% confidence intervals at a centre frequency of $78\pm 1$~MHz or $z\simeq17$ \cite{Bowman:2018yin}. To resolve this discrepancy, either one has to increase the background radio radiation or decrease the gas temperature. Both the scenarios have been explored in several literatures   \cite{Ewall-Wice2018, Jana:2018, Feng2018, Lawson:2019,Natwariya:2021, Lawson:2013, Levkov:2020, Natwariya:2020, Brandenberger:2019, Chianese:2019,Bhatt2019pac, Tashiro:2014tsa, Barkana:2018lgd, SIKIVIE2019100289, Mirocha2019, Ghara2019}. However, increasing the background radio radiation or cooling the IGM gas by the non-standard mechanisms are not well known and are debatable issues  \cite{Munoz2018, FRASER2018159, Bransden1958, Barkana:2018nd, Berlin2018, Kovetz2018, Munoz2018a, Slatyer2018, Amico:2018, Mitridate:2018}.  Therefore, we have not considered any methods to increase the background radiation above CMB or cooling of the IGM gas.  In this work, we study  projected bounds on spinning PBHs such that 21-cm differential brightness temperature does not change more than a factor of 1/4 from $\Lambda$CDM framework based theoretical prediction.

The fraction of DM in the form of PBHs is constrained from various astrophysical observations and theoretical predictions. PBHs with mass smaller than $\sim\mathcal{O}(10^{15}~{\rm g})$ may have evaporated as of now and can be constrained from the impact on big bang nucleosynthesis by evaporated particles, background radiation etc. Higher mass PBHs can be constrained by the effect on large-scale structures, gravitational wave and lensing, and impact on thermal and ionization history of the IGM (for details, see the recent reviews \cite{carr:2021, Green:2021, Carr:2020} and references therein). In the context of the 21-cm signal, the upper bound on the $f_{\rm PBH}$ can be found in Refs. \cite{Hektor:2018, Clark:2018, Mena:2019, Yang:2020, Halder_2021, Tashiro_2021, Yang_2020, Pablo:2021}. Angular momentum is a fundamental property of a black hole, and it can modify the Hawking evaporation drastically. In the case of rotating PBHs, authors of the Refs. \cite{Dasgupta:2020, Laha:2021} have reported the various types of bound on $f_{\rm PBH} $ as a function of PBHs mass and spin. Future collaboration, All-sky Medium Energy Gamma-ray Observatory (AMEGO)\footnote{\href{https://asd.gsfc.nasa.gov/amego/index.html}{https://asd.gsfc.nasa.gov/amego/index.html}} will be able to constrain some parameter space for the rotating  PBHs \cite{Ray:2021}.

%=========================================================================

\section{Thermal History Of IGM}
A rotating black hole with angular momentum $J_{\rm PBH}$ and having mass $M_{\rm PBH}$ can be defined with a rotation parameter, $a_*=J_{\rm PBH}/(G\,M_{\rm PBH}^2)$ \cite{Page::1976}, where $G$ is the gravitational constant. Black holes can get their spin depending on generation mechanisms, merger or accretion \cite{Kesden_2010, Cotner_2017, Harada_2021, Luca_2019, Luca_2020, Harada_2017, K_hnel_2020, flores:2021, Arbey_2020, He_2019, Cotner_2019}. PBHs with higher mass can have a lifetime larger/comparable than the age of the Universe. Therefore, they have enough time to accrete  mass and spin up \cite{Dong_2016}. Rotating black hole with higher spin ($a_*\rightarrow1$) injects more energy into IGM and evaporates faster than non-rotating BHs \cite{Chandrasekhar:1977, Taylor_1998, Arbey::2020, ArbeyAJ:2021}. Therefore, we expect that the bounds on $f_{\rm PBH}$ to be more stringent compared to non-rotating PBHs. The energy injection per unit volume per unit time due to $e^{\pm}$ and photons into IGM, for monochromatic mass distribution of PBHs, can be written as \cite{Laha:2021, mittal:2021},
{\small{
		\begin{alignat}{2}
		\Gamma_{\rm PBH}^{e^{\pm}}(z,a_*)&=2 \int\left[ f_c^e(E-m_e,z)\,(E-m_e)\left(\frac{d^2 N_{e}}{dt\,dE}\right)\, \right]\,n_{\rm PBH}\, dE\,,\label{eq1}\\
		\Gamma_{\rm PBH}^{\gamma}(z,a_*)&=\int\left[\ f_c^\gamma(E,z)\,E\,\left(\frac{d^2 N_\gamma}{dt\,dE}\right)\ \right]\,n_{\rm PBH}\ dE\,.\label{eq2}
		\end{alignat}}}
Energy injection into IGM happens by three processes: heating, ionization, and excitation of the gas \cite{Slatyer:2016, Slatyer::2016, Liu_2020}. $f_c^i$ represents the energy deposition efficiency into IGM. Here, $c$ stands for above-mentioned three channels and $i\equiv({\rm electron/positron,\, photon})$ stands for different types of injected particles. The factor of 2 in equation \eqref{eq1} accounts for the total contribution of electrons and positrons. $n_{\rm PBH}=f_{\rm PBH}\,(\rho_{\rm DM}/M_{\rm PBH})$ is the number density of the PBHs, and $\rho_{\rm DM}$ is the dark matter energy density. $d^2 N^i/(dt\,dE)\equiv d^2 N^i/(dt\,dE)\,\big(E,M_{\rm PBH}, a_*\big)$ represents the number of particles emitted by black hole per unit time per unit energy \cite{Page::1976, Mac:1990, Laha:2021, Arbey_2019}. We use the \texttt{BlackHawk} code\footnote{\href{https://blackhawk.hepforge.org/}{https://blackhawk.hepforge.org/}} to calculate the spectra due to photons and electrons/positrons; we take both the primary and secondary Hawking evaporation spectra into account \cite{Arbey_2019, ArbeyJ:2021}.
%==========================================================================

In the presence of Hawking radiation, the thermal evolution of the gas \cite{Chluba2015, Amico:2018},
\begin{alignat}{2}
\frac{dT_{\rm gas}}{dz}=2\,\frac{T_{\rm gas}}{1+z}+\frac{\Gamma_c}{(1+z)\,H}(T_{\rm gas}-T_{\rm CMB})-\frac{2\ \,\Gamma_{\rm PBH}\,}{3\,N_{\rm tot}(1+z)\,H}\,,\label{eq3}
\end{alignat}
here, $\Gamma_{\rm PBH}=\Gamma_{\rm PBH}^{e^{\pm}}+\Gamma_{\rm PBH}^{\gamma}$ is the total energy injection per unit time and per unit volume into IGM, $N_{\rm tot}=N_{\rm H}\,(1+f_{\rm He}+X_e)$ is the total number density of the gas, $N_{\rm H}$ is the hydrogen number density,  $f_{\rm He}=N_{\rm He}/N_{\rm H}$, $N_{\rm He}$ is the helium number density, $X_e=N_e/N_{\rm H}$ is the free electron fraction and $N_e$ is the free electron number density. $\Gamma_c$ stands for the Compton scattering rate \cite{Schleicher:2008aa,Natwariya:2020}. We consider the following numerical values of the cosmological parameters: $h=0.674$, $\Omega_{\rm M }=0.315$, $\Omega_{\rm b}=0.049$ and $T_{\rm CMB}|_{z=0}=2.725$~K \cite{Planck:2018,Fixsen_2009}. To compute the energy deposition efficiency, thermal and ionization history of the Universe, we use \texttt{DarkHistory}\footnote{\href{https://darkhistory.readthedocs.io/en/master/}{https://darkhistory.readthedocs.io/en/master/}} package with necessary modifications \cite{Liu_2020}. 
%

%===========================================================================

\begin{figure*}
	\begin{center}
		\subfloat[] {\includegraphics[width=3.4in,height=2.2in]{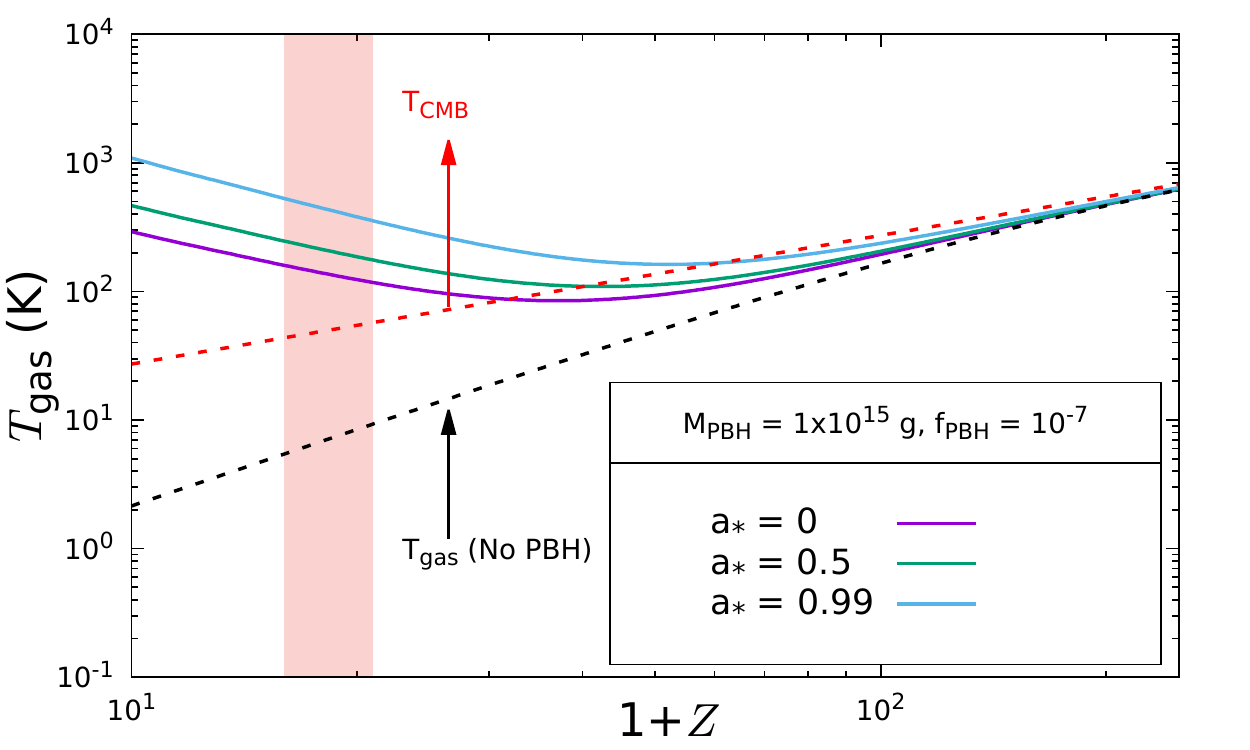}\label{plot:1a}}
		\subfloat[] {\qquad\includegraphics[width=3.4in,height=2.2in]{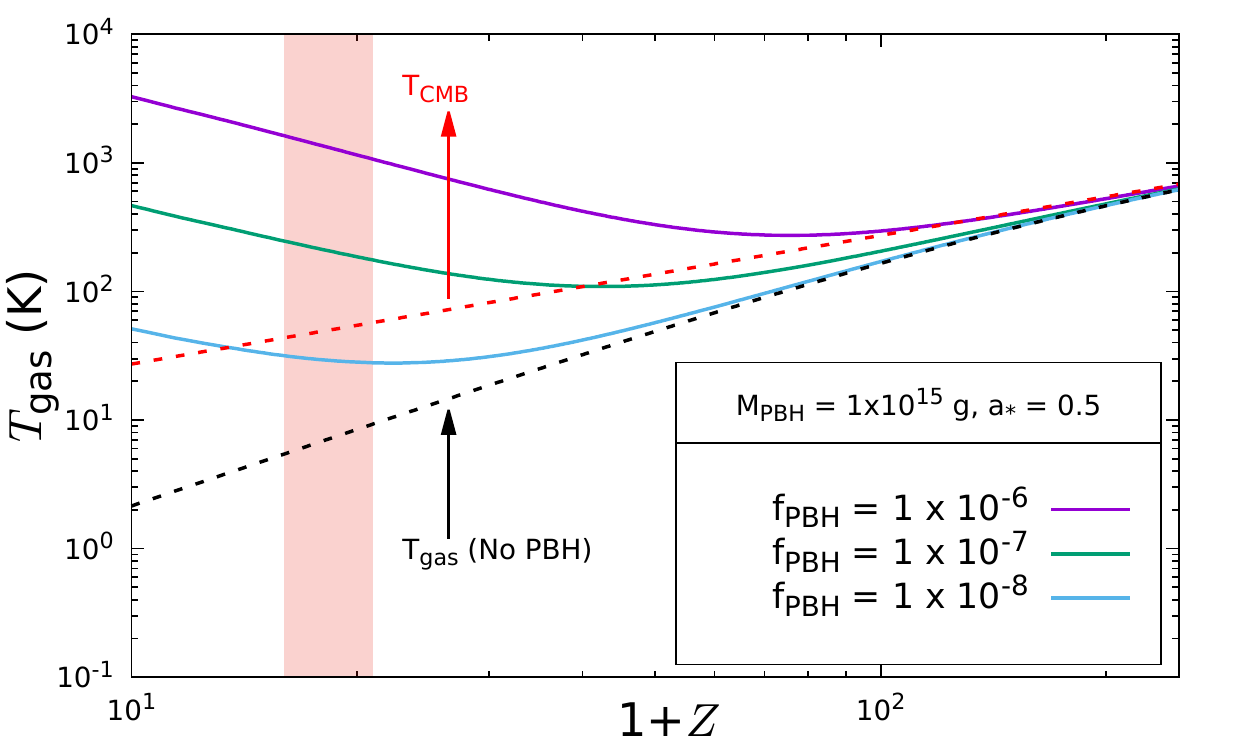}\label{plot:1b}}  
	\end{center}
	\caption{ The gas temperature evolution with redshift for evaporating primordial black hole. The red dashed lines represent the CMB temperature evolution. The black dashed lines depicts the $T_{\rm gas}$ when there is no PBHs. The shaded region corresponds to the redshift $15\leq z \leq 20$ (EDGES observed signal). In plot \eqref{plot:1a}, we consider PBHs mass and $f_{\rm PBH}$ to $1\times10^{15}$~g and $10^{-7}$, respectively, and vary the spin of PBHs.  In plot \eqref{plot:1b}, we keep $M_{\rm PBH}=1\times10^{15}$~g and $a_*=0.5$ constant and vary $f_{\rm PBH}$.}\label{plot:1_1}
\end{figure*}

\begin{figure}
	\begin{center}
		{\includegraphics[width=3.4in,height=2.2in]{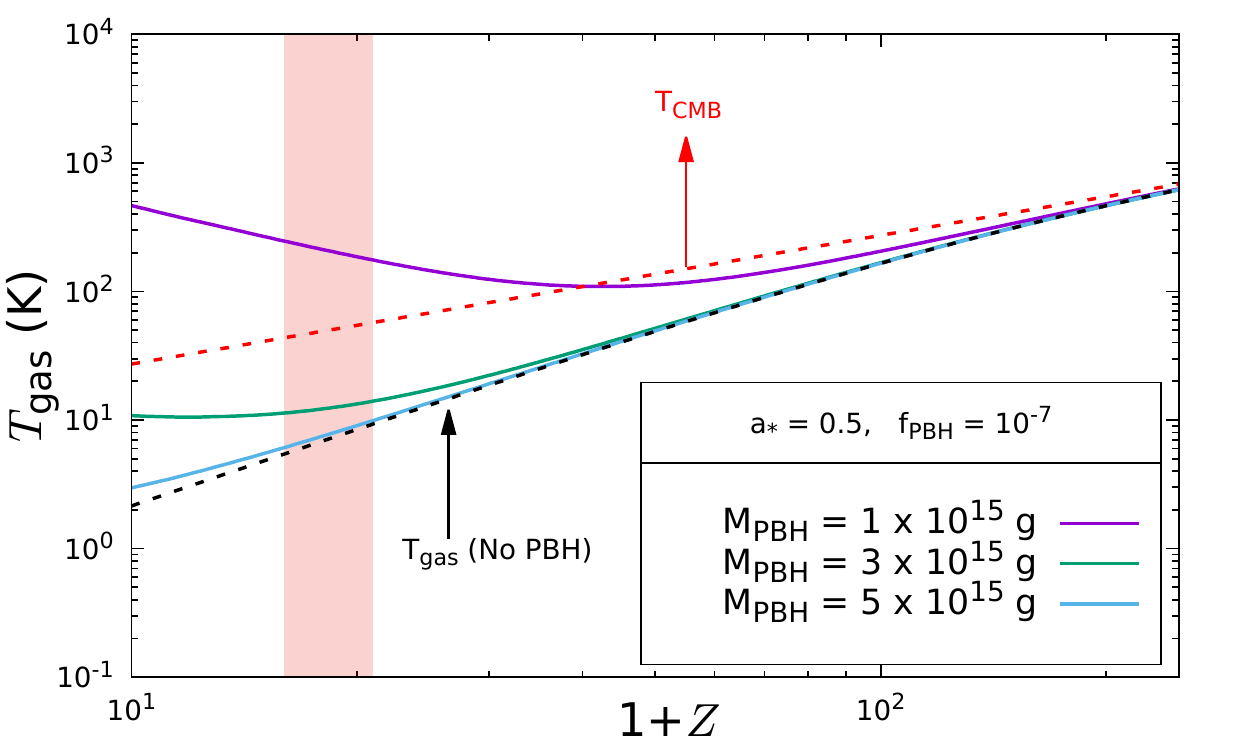}\label{plot:1c}} 
	\end{center}
	\caption{ The caption is the same as in Figure \eqref{plot:1_1}, except, here, we vary the mass of PBHs and keep spin and $f_{\rm PBH}$ to $0.5$ and $10^{-7}$, respectively.}\label{plot:1_2}
\end{figure}

%===========================================================================

\section{Results and Discussion}   
Following the Refs. \cite{Mesinger:2007S, Mesinger:2011FS, Pritchard_2012, Mittal:2020}, we write the global 21-cm differential brightness temperature as,  \begin{alignat}{2}
T_{21}=27\,X_{\rm HI}\,  \left(1-\frac{T_{\rm R}}{T_{\rm S}}\right)\,\left(\frac{0.15}{\Omega_{\rm m }}\,\frac{1+z}{10}\right)^{1/2}\left(\frac{\Omega_{\rm b}h}{0.023}\right)~{\rm mK}\,,\label{2}
\end{alignat}
here, $X_{\rm HI}=N_{\rm HI}/N_{\rm H}$ is the fraction of neutral hydrogen in the Universe, and $N_{\rm HI}$ is the  neutral hydrogen number density. $T_{\rm S}$ is the spin temperature, and it is characterized by the number density ratio of $1S$ triplet and singlet hyperfine states of the neutral hydrogen atom. In the cosmological scenarios, there are mainly three processes that can affect the spin temperature: background radio radiation, Ly$\alpha$ radiation from the first stars and collisions of a hydrogen atom with another hydrogen atom, residual electron or proton. In the detailed balance between the population of $1S$ singlet and triplet state, one can write the spin temperature as \cite{Field, Pritchard_2012},
\begin{alignat}{2}
T_{\rm S}^{-1}=\frac{T_{\rm R}^{-1}+x_\alpha\,T_{\alpha}^{-1}+x_c\,T_{\rm gas}^{-1}}{1+x_\alpha+x_c}\,,
\end{alignat}
here, $T_\alpha$ is the Ly$\alpha$ colour temperature. $x_\alpha$ is the Ly$\alpha$ coupling coefficient due to Wouthuysen-Field effect \cite{1952AJ.....57R..31W, Field}. $x_c$ is the collisional coupling coefficient due to scattering  between hydrogen atoms or scattering of hydrogen atoms with other species such as electrons and protons (\cite{Pritchard_2012} and references therein). The colour temperature can be taken as gas temperature, $T_\alpha\simeq T_{\rm gas}$, due to the repeated scattering between Ly$\alpha$ photons and gas \cite{Field, 1959ApJ...129..536F, Pritchard_2012}. After the formation of the first stars ($z\sim30$), their Ly$\alpha$ radiation causes the hyperfine transition in the neutral hydrogen atom, and the $x_\alpha$ starts dominating over other couplings \cite{Pritchard_2012}. Therefore, at the redshift 17.2, spin temperature can be approximated as $T_{\rm S}\simeq T_{\rm gas}$, when the background radiation temperature $T_{\rm R}=T_{\rm CMB}$ \cite{Pritchard_2012, Natwariya:2021}.  In the standard cases, the background radiation contribution is assumed solely by CMB radiation, $T_{\rm R}=T_{\rm CMB}$.  In the present work, we do not consider X-ray heating of the gas due to the uncertainty of the known physics of the first stars. The inclusion of X-ray heating will further strengthen our projected constraints. Here, it is to be noted that the gas temperature may increase due to the energy transfer from the background radiation to the thermal motions of the gas mediated by Ly$\alpha$ radiation from the first stars \cite{Venumadhav:2018}. However, due to the uncertainty in known physics of the first star formation, we do not include this effect also. The inclusion of this effect will further strengthen our  projected bounds on $f_{\rm PBH}$. Depending on the ratio $T_{\rm CMB}/T_{\rm S}$, there can be three scenarios: absorption ($T_{\rm CMB}>T_{\rm S}$), emission ($T_{\rm CMB}<T_{\rm S}$) or no signal ($T_{\rm CMB}=T_{\rm S}$). At redshift $17.2$, to get $T_{21}\leq-150$~mK, we require $T_{\rm gas}\leq9.62$~K. Here, $X_{\rm HI}\simeq1-X_e$, and in our case at required redshift we get $X_e\lesssim\mathcal O (10^{-3})$. Therefore, $X_{\rm HI}$ can be regarded as unity. 

%=========================================================================
%
\begin{figure*}
	\begin{center}
		\subfloat[] {\includegraphics[width=3.3in,height=3.1in]{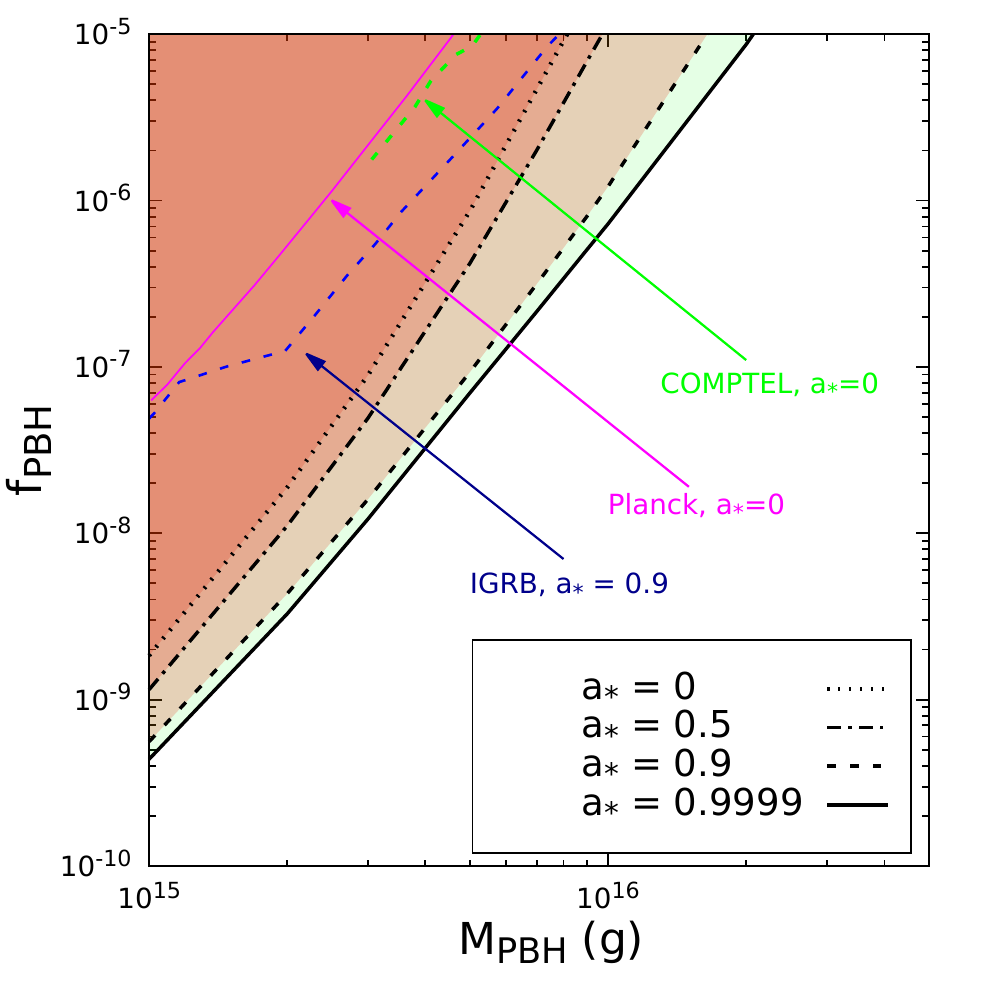} \qquad \label{plot:2a}}
		\subfloat[] {\qquad \includegraphics[width=3.3in,height=3.1in]{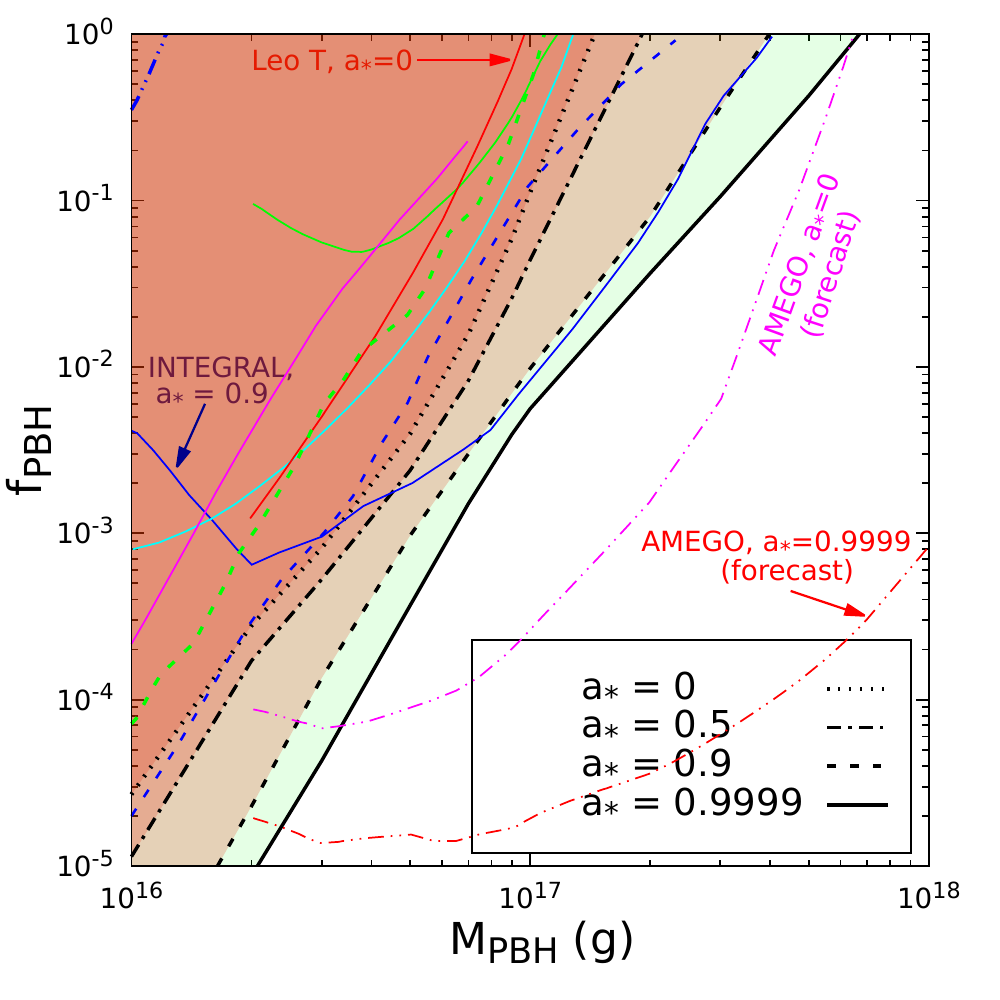}\label{plot:2b}}  
	\end{center}
	\caption{The upper projected bounds on the dark fraction of matter in the form PBHs ($f_{\rm PBH}=\Omega_{\rm PBH}/\Omega_{\rm DM}$) as a function of PBHs mass for varying spin ($a_*$) of PBHs. The shaded regions are excluded from our analysis for $f_{\rm PBH}$ when $a_*=0$ (dotted black line), 0.5 (dot-dashed black line), 0.9 (dashed black line) and 0.9999 (solid black line). The dashed blue curve depicts the upper constraint on $f_{\rm PBH}$ by observations of the diffuse Isotropic Gamma-Ray Background (IGRB) for $a_*=0.9$ \cite{Arbey::2020}. The double-dot-dashed blue curve represents the upper constraint on $f_{\rm PBH}$ from Diffuse Supernova Neutrino Background (DSNB) searches at Super-Kamiokande, while the solid blue line represents the INTErnational Gamma-Ray Astrophysical Laboratory (INTEGRAL) observation of 511~KeV $\gamma$-ray lines at Galactic centre constraint on $f_{\rm PBH}$ for $a_*=0.9$ \cite{Dasgupta:2020}.  The double-dot-dashed magenta (red) line represents the AMEGO forecast for $a_*= 0\ (a_*=0.9999)$ \cite{Ray:2021}. Near future, AMEGO collaboration will be able to probe the parameter-space above the magenta (red) double-dot-dashed curve for  $a_*= 0\ (a_*=0.9999)$. The solid green line stands for 95\% confidence level bound from INTEGRAL observation of Galactic gamma-ray flux for non-spinning PBHs \cite{Laha:2020}. Solid cyan curve depicts the upper bound from observing the 511 KeV $\gamma$-ray lines at the Galactic centre by assuming all the PBHs within a 3 Kpc radius of the Galactic centre for non-spinning PBHs \cite{Laha:2019}. The magenta solid line represents the Planck constraint \cite{Clark:2017}. The red solid line depicts the dwarf galaxy Leo T constraint \cite{Kim_2021} and the green dashed line shows the COMPTEL bound \cite{Coogan:2021} for non-spinning PBHs. }\label{plot:2}
\end{figure*}
%
%========================================================================
% 

In Figures \eqref{plot:1_1} and \eqref{plot:1_2}, we present IGM gas temperature evolution as a function of redshift for different PBH masses, spins and fractions of DM in the form of PBHs. The shaded region corresponds redshift range, $15-20\,$.  The red dashed curves in all plots depict the CMB temperature evolution, while the black dashed line represents the gas temperature when there are no evaporating PBHs. In Figure \eqref{plot:1a}, we keep mass and $f_{\rm PBH}$ to $1\times10^{15}$~g and $10^{-7}$, respectively, and vary the spin of PBHs. As expected, when we increase the spin of PBHs, the gas temperature rises significantly in the shaded region. The solid violet curve represents the case when the spin of PBHs is 0. Increasing the spin to 0.5 (solid green line), the gas temperature increases. Continuously increasing spin to 0.99 (solid cyan line), the gas temperature rises further. In Figure \eqref{plot:1b}, we keep  $M_{\rm PBH}=1\times10^{15}$~g, spin to 0.5 and vary $f_{\rm PBH}$. In this plot, as we increase the $f_{\rm PBH}$ from $10^{-8}$ (solid cyan line) to $10^{-6}$ (solid violet line), the IGM heating rises rapidly. If the gas temperature becomes larger than the CMB temperature in the shaded region, it can erase the 21 cm absorption signal; instead, it may give an emission signal. Therefore, at desired redshift (in our scenario $z = 17.2$), one has to keep $T_{\rm gas}< T_{\rm CMB}$ to get an absorption signal. Increasing $f_{\rm PBH}$, for a given mass, the number density of PBHs increases, resulting in more energy injection into IGM by PBHs Hawking evaporation. Therefore, increasing the $f_{\rm PBH}$, the gas temperature rises. In Figure \eqref{plot:1_2},  we vary the mass of PBHs and keep spin and $f_{\rm PBH}$ constants to $0.5$ and $10^{-7}$, respectively. In this plot, as we increase the mass of PBHs from $1\times10^{15}$~g (solid violet line) to $5\times10^{15}$~g (solid cyan line), the gas temperature decreases. It happens for two reasons:  (i) Increasing the mass of PBHs leads to a decrease in the  total power contributions from Hawking evaporation of PBHs \cite{Mac:1990}. (ii)  Ignoring the integral dependency in equations \eqref{eq1} and \eqref{eq2}, $\Gamma_{\rm PBH}^{e^\pm}$ and $\Gamma_{\rm PBH}^{\gamma}$ are proportional to  $n_{\rm PBH}=f_{\rm PBH}\,(\rho_{\rm DM}/M_{\rm PBH})$. For a fixed dark-matter energy density and $f_{\rm PBH}$, the number density of PBHs increases by decreasing the black hole mass. Thus, energy injection into IGM per unit volume and time ($\Gamma_{\rm PBH}$) increases, and one gets more heating of the gas.

In Figure \eqref{plot:2}, we plot the upper  projected bounds on the fraction of DM in the form of PBHs as a function of PBHs mass for different spins. Here, we have considered that 21-cm differential brightness temperature, $T_{21}$, remains $-150$~mK at redshift $z=17.2$. We vary the mass of PBHs from $10^{15}$~g to $10^{18}$~g. The shaded regions in both the plots are excluded for the corresponding PBH spins. The dashed blue curve represents the upper constraint on $f_{\rm PBH}$ by observations of the diffuse Isotropic Gamma-Ray Background (IGRB)  \cite{Arbey::2020}. The double-dot-dashed blue curve represents the upper constraint on $f_{\rm PBH}$ from Diffuse Supernova Neutrino Background (DSNB) searches at Super-Kamiokande, while the solid blue line represents the INTErnational Gamma-Ray Astrophysical Laboratory (INTEGRAL) observation of 511 KeV $\gamma$-ray line at Galactic centre constraint on $f_{\rm PBH}$ for $a_*=0.9$ \cite{Dasgupta:2020}.  For $a_*=0$, the observation at the Jiangmen Underground Neutrino Observatory (JUNO) will be able to place a 20 times stronger bound on the upper allowed value of $f_{\rm PBH}$ for $ M_{\rm PBH}=10^{15}$~g compared to Super-Kamiokande \cite{Wang:2021, Dasgupta:2020}. The double-dot-dashed magenta (red) line represents the AMEGO forecast for $a_*= 0\ (a_*=0.9999)$ \cite{Ray:2021}. In the near future, AMEGO collaboration will be able to probe the parameter-space above the magenta (red) double-dot-dashed curve for  $a_*= 0\ (a_*=0.9999)$. Solid green line stands for  95\% confidence level bound from INTEGRAL observation of Galactic $\gamma$-ray flux for non-spinning PBHs \cite{Laha:2020}. The solid cyan curve depicts the upper bound from the observation of 511 KeV $\gamma$-ray lines at the Galactic centre by assuming all the PBHs within a 3 Kpc radius of the Galactic centre for non-spinning PBHs \cite{Laha:2019}.  For the comparison, we have also plotted the bounds from Planck \cite{Clark:2017}, Leo T \cite{Kim_2021} and COMPTEL \cite{Coogan:2021} observations for non-spinning PBHs. In Figure \eqref{plot:2a}, $f_{\rm PBH}$ varies from $1\times10^{-10}$ to $1\times10^{-5}$, while, in Figure \eqref{plot:2b}, it varies from $1\times10^{-5}$ to its maximum allowed value 1 ($\Omega_{\rm PBH}=\Omega_{\rm DM}$). In Figure \eqref{plot:2}, as we increase the value of spin from $0$ to its extremal value, $0.9999$, the upper bounds become more stringent. This is due to an increment in  evaporation of PBHs, and it results in more energy injection into the IGM \cite{Page::1976, Page:::1976, Page:1977}. As discussed earlier, increasing the mass of PBHs, energy injection into IGM decreases. Subsequently, one gets more window to increase the gas temperature or $f_{\rm PBH}$, and the upper bound becomes weaker. Therefore, in Figure \eqref{plot:2}, the upper bound on $f_{\rm PBH}$ weakens as we increase the mass. Our upper projected constraint on $f_{\rm PBH}$ for $a_*=0.9$ is comparable to the INTEGRAL observation of 511~KeV $\gamma$-ray lines for PBHs mass larger than $\sim8\times10^{16}$ and becomes stronger for smaller PBH masses. Also, compared to IGRB \cite{Arbey::2020} and DSNB \cite{Dasgupta:2020}, our  projected bounds are stringent for the considered mass range of PBHs. We find the most robust lower projected constraint on the mass of PBHs, which is allowed to constitute the entire dark matter,  to $1.5\times10^{17}$~g, $1.9\times10^{17}$~g, $3.9\times10^{17}$~g and $6.7\times10^{17}$~g for PBH spins 0, 0.5, 0.9 and 0.9999, respectively. The lower bound on $M_{\rm PBH}$ for $\Omega_{\rm PBH}=\Omega_{\rm DM}$, for extremal spinning PBHs is nearly four times larger than non-spinning PBHs.

%============================================================================================

\section{Conclusions}
Spinning primordial black holes can substantially affect the ionization and thermal history of the Universe.  Subsequently, it can modify the 21-cm absorption signal in the cosmic dawn era by injecting energy due to Hawking evaporation. We study the upper  projected bounds on the fraction of dark matter in the form of PBHs as a function of mass and spin, considering that the 21-cm differential brightness temperature does not change more than a factor of 1/4 from the theoretical prediction based on the $\Lambda$CDM framework.  Our  projected constraints are stringent compared to DSNB, INTEGRAL observation of the 511~KeV line, IGRB, Planck, Leo T and COMPTEL. In the near future, AMEGO collaboration will be able to probe some parameter space in our considered mass range of PBHs. In the present work, we have considered the monochromatic mass distribution of PBHs. The allowed parameter space can also be explored for different PBHs mass distributions such as log-normal, power-law, critical collapse, etc. \cite{Arbey_2019}.  Here, it is to be noted that we have not considered heating of IGM gas due to X-ray from the first stars in the vague of known physics of the first stars. The inclusion of X-ray heating will further strengthen our  projected bounds.

\vspace{1cm}
%============================================================================================
\section{Acknowledgements}

The authors would like to acknowledge Prof. Jitesh R Bhatt and Ranjan Laha for valuable comments and suggestions, and the TEQIP-III sponsored Workshop on Astroparticle Physics and Cosmology at the National Institute of Technology Meghalaya. We thank Alexandre Arbey and Jérémy Auffinger for providing the new version of the BlackHawk code in advance. T. S. would like to acknowledge the support from the Dr. D. S. Kothari Postdoctoral fellowship scheme No. F.4-2/2006 (BSR)/PH/20-21/0163 Finally, the authors would like to thank the Referee for the suggestions and a detailed report that significantly improved the quality of the manuscript.   
%=============================================================================


\end{document}